\documentstyle[prl,aps,preprint]{revtex}
\begin{document}
\newcommand{\Z}{{\mbox{\rm Z}}}
\newcommand{\R}{{\mbox{\rm R}}}
\newcommand{\C}{{\mbox{\rm C}}}
\newcommand{\dd}{{\mbox{\rm d}}}
\setcounter{page}{0}
\def\footnoterule{\kern-3pt \hrule width\hsize \kern3pt}
\tighten
\title{REPRESENTATION OF COMPLEX PROBABILITIES\thanks
{This work is supported in part by funds provided by the U.S.
Department of Energy (D.O.E.) under cooperative 
research agreement \#DF-FC02-94ER40818 and Spanish DGICYT grant
no. PB92-0927.}} 

\author{L.L. Salcedo\footnote{Email address: {\tt salcedo@goliat.ugr.es}}}
\address{Center for Theoretical Physics \\
Laboratory for Nuclear Science \\
and Department of Physics \\
Massachusetts Institute of Technology \\
Cambridge, Massachusetts 02139, U.S.A. \\
and \\
Departamento de F\'{\i}sica Moderna \\
Universidad de Granada \\
E-18071 Granada, Spain \\
{~}}

\date{MIT-CTP-2550,~  {~~~~~} July 1996}
\maketitle

\thispagestyle{empty}

\begin{abstract}

Let a ``complex probability'' be a normalizable complex distribution
$P(x)$ defined on $\R^D$. A real and positive probability distribution
$p(z)$, defined on the complex plane $\C^D$, is said to be a positive
representation of $P(x)$ if $\langle Q(x)\rangle_P = \langle
Q(z)\rangle_p$, where $Q(x)$ is any polynomial in $\R^D$ and $Q(z)$ its
analytical extension on $\C^D$. In this paper it is shown that every
complex probability admits a real representation and a constructive
method is given.  Among other results, explicit positive
representations, in any number of dimensions, are given for any complex
distribution of the form Gaussian times polynomial, for any complex
distributions with support at one point and for any periodic Gaussian
times polynomial.

\end{abstract}


\pacs{PACS 11.15.Ha 02.70.Lq 02.60.Cb 02.50.Ey}
\eject

\section{Introduction}

In quantum physics there are instances of averages where the role of
probability distribution is played by a distribution taking complex
values. Consider the functional integral formulation of field theory
\cite{Ra90}. There, the time ordered expectation value of observables
takes de form $\langle T{\cal O}[\phi]\rangle = N\int{\cal
D}\phi(x)\,e^{iS[\phi]}{\cal O}[\phi]$, where $S[\phi]$ is the action
functional and $N$ a normalization constant. This is a first instance
of a ``complex probability distribution'', namely, the Boltzmann
weight $P[\phi]=Ne^{iS[\phi]}$. In the continuum, such functional
integral is not sufficiently well-behaved and only its Euclidean
version can be given a rigorous meaning \cite{GJ87}. Within a lattice
regularization, the Minkowski version is mathematically well-defined,
nevertheless the Wick rotation is performed in this case too. This is
because, in most cases, in the Euclidean theory the Boltzmann weight
becomes a real and positive probability distribution. This is
important in practice since straightforward Monte Carlo is only
defined for positive probabilities. There are cases, however, when
even Euclidean field theory presents complex actions. Indeed, the
statistical interpretation of the quantum theory requires the
Boltzmann weight to be reflection positive, but not directly positive
\cite{OS73}. Instances of complex Euclidean actions occur after
integration of fermions, since the fermionic determinant is not
positive definite; if there are non vanishing chemical potentials; in
gauge theories in the presence of Wilson loops or topological
$\theta$-terms or in general after inserting projection operators in
the path integral to select particular sectors of the theory
~\cite{FN83,BD84,AF86,AY86}.  Also, two dimensional fermions can be
brought to a bosonic complex action form~\cite{KL92}.

As we have said, the computation of averages in
the presence of a complex probability distribution
poses a practical problem, namely, the Monte Carlo method cannot be
used directly to sample the probability since this method only makes
sense for true, i.e. real and positive, probabilities.
The standard approach to complex probabilities in numerical
simulations~\cite{FN83,BD84} is to factorize a real and positive part
to be used as input for some Monte Carlo method and include the
remainder in the observable. That is, if the complex probability is
$P(x)=P_0(x)F(x)$ with $P_0(x)$ positive, the expectation values can
be obtained as
\begin{equation}
\langle{\cal O}(x)\rangle_P
=\frac{\langle{\cal O}(x)F(x)\rangle_{P_0}}{\langle F(x)\rangle_{P_0}}\,.
\end{equation}
Of course, the same formula can be used when $P(x)$ itself is
positive. The problem with this approach is that it violates the
importance sample principle, since we are not sampling the true
probability and that increases the dispersion of Monte Carlo data. For
instance, $\langle F(x)\rangle_{P_0}$ may be small, thereby
introducing large error bars.

An alternative approach is to look for a positive probability $p(z)$
in the complex configuration space which gives the same
expectation values as $P(x)$, i.e., $\langle{\cal O}(x)\rangle_P =
\langle{\cal O}(z)\rangle_p$, where ${\cal O}(z)$ is the analytical
extension of ${\cal O}(x)$.  The usual way of constructing such a
probability is by means of the complex Langevin algorithm
\cite{Kl83,Pa83}. In this approach the configuration is updated through a
standard Langevin algorithm with the complex action. Since the
drift term is complex, the complex extension of the configuration
space is sampled as well. Whenever the random walk possesses an
equilibrium configuration, it is sampling the complex configuration
space with a real and positive probability distribution $p(z)$. We
have then traded a complex probability $P(x)$ on $\R^D$ by a positive
probability $p(z)$ on $\C^D$. If $p(z)$ happens to be equivalent to
$P(x)$ in the sense of expectation values, we have succeeded in
sampling the complex probability. Successful implementations of the
algorithm have been obtained in some practical cases, such as
two dimensional compact QED with static charges~\cite{AF85}.
In general, however, the complex Langevin
algorithm poses two problems. First, it not always converges to an
equilibrium distribution. Second and more subtle, for some actions it
seems to converge to an equilibrium distribution which is not
equivalent to the original complex probability~\cite{AF86,FO86,Sa93}, (see
however~\cite{Se94}). 
Such phenomenon has been found in practically relevant cases such
as QCD with a Wilson loop~\cite{AF86,FO86,FO94}.

In the present paper we consider the problem of constructing a
positive representation directly, independently of the Langevin
algorithm. Several properties of representations of complex
probabilities on $\R^D$ by probabilities on $\C^D$ are noted. A
constructive method is given to obtain real (although not necessarily
positive) representations of very general complex probabilities.
Positive representations are explicitly constructed
for some probabilities which
are beyond the present applicability of the complex Langevin
algorithm. These include Gaussian times polynomial, distributions
with support at one point, and periodic Gaussian times polynomial.
In all cases, such representations are not unique.

These results are of great interest from the point of view of
applications. This is not because the constructions found here are of
direct usefulness to carry out numerical calculations; there are far
more natural ways to compute expectations values with complex Gaussian
times polynomial distributions. The interest lies in the
following. The negative results found up to now with the complex
Langevin algorithm in some systems would make one to have reasonable
doubts of whether a positive representation exists at all for those
systems. Moreover, the momenta of any positive probability on $\C^D$
are bounded to satisfy some inequalities among them. It might happen
that those bounds were incompatible with the momenta of the given
complex probability on $\R^D$ in some cases.  At present, the
necessary and sufficient conditions for a positive representation to
exist are not known. The results of this paper suggest, however, that
such representation exists quite generally since the set of Gaussian
times polynomial is dense in $L^2(\R^D)$. Our results tend to support
the idea that there is no obstruction of principle for positive
representations to exits. This is the main insight of this work.

\section{Representation of complex probabilities}

The complex probabilities $P(x)$ to be considered here will be
tempered distributions on $\R^D$ of a restricted class, namely, those
which are the inverse Fourier transform of an ordinary function ${\tilde
P}(k)$ (locally integrable and at most of polynomial growth at
infinity), with ${\tilde P}(k)$ non vanishing at the origin and
analytical at that point. These conditions allow for a natural
definition of $\int x_{i_1}\cdots x_{i_n}P(x)\dd^Dx$ through the
Taylor expansion of ${\tilde P}(k)$ at $k=0$. In particular
$\int P(x)\dd^Dx$ will be non vanishing. The expectation value 
associated to $P(x)$ is defined for any polynomial $Q(x)$ as
\begin{equation}
\langle Q(x)\rangle_P =
\frac{\int Q(x)P(x)\dd^Dx}{\int P(x)\dd^Dx}\,.
\label{eq1}
\end{equation}
Likewise, we can consider complex probabilities on $\C^D$ as the 
class of distributions defined above on $\R^{2D}$.
For any such distribution, $p(z)$, the
expectation value takes the form
\begin{equation}
\langle q(z)\rangle_p =
\frac{\int q(z)p(z)\dd^{2D}z}{\int p(z)\dd^{2D}z}\,.
\label{eq2}
\end{equation}
where $z_j=x_j+iy_j$, $\dd^{2D}z=\dd^Dx\dd^Dy$ and $q(z)$ is an arbitrary
polynomial of $z$ and its complex conjugate $z^*$.

By definition, $p(z)$ is a representation of $P(x)$ 
if $\langle Q(x)\rangle_P = \langle Q(z) \rangle_p$, where $Q(x)$ is
any polynomial on $\R^D$ and $Q(z)$ its analytical extension on
$\C^D$. Equivalently, one can demand
$\langle x_{i_1}\cdots x_{i_n}\rangle_P= \langle z_{i_1}\cdots
z_{i_n}\rangle_p$ for any set of indices, where $i_r=1,\dots,D$ and
$n=0,1,2,\dots$. Two complex probabilities on $\C^D$ will be called
equivalent if they have the same expectation values on every
analytical polynomial. In general, two equivalent
probabilities will not coincide on expectation values of non
analytical polynomials $\langle z_{i_1}\cdots z_{i_n}z^*_{j_1}\cdots
z^*_{j_m}\rangle$. A representation will be called real if $p(z)$ is
real, positive if $p(z)$ is non negative and unitary if
$\int p(z)\dd^{2D}z=\int P(x)\dd^Dx$.  Our goal is then to find
positive representations of complex probabilities.

We will proceed by noting different ways to obtain new representations
from known ones. A first obvious way is by means of complex affine
transformations. Let $A$ be a non singular complex $D\times D$ matrix,
and $a\in \C^D$, and assume that $P_0(z)$ is an analytical function in
a region including $\R^D$ and $A\R^D+a$ such that $P_0(x)$ and
$P(x)=\det(A)P_0(Ax+a)$ are both complex probabilities. Then if
$p_0(z)$ is a unitary representation of $P_0(x)$ so is
$p(z)=|\det(A)|^2p_0(Az+a)$ of $P(x)$: for any polynomial $Q(x)$
\begin{eqnarray}
|\det(A)|^2 && \int Q(z) p_0(Az+a) \dd^{2D}z = 
\int Q(A^{-1}(z-a)) p_0(z) \dd^{2D}z  \nonumber\\
&&= \int Q(A^{-1}(x-a)) P_0(x) \dd^Dx =
\det(A) \int Q(x) P_0(Ax+a) \dd^Dx \,.
\end{eqnarray}
Furthermore, $p(z)$ is positive if $p_0(z)$ is positive. Another
construction follows from linear combination. If $p_i(z)$ are unitary
representations of $P_i(x)$, so is $p(z)=\sum_{i=1}^nb_ip_i(z)$ of
$P(x)=\sum_{i=1}^nb_iP_i(x)$. Again, if $p_i(z)$ are positive and
$b_i$ non negative, $p(z)$ is positive too.

Let us define the partial derivatives $\partial_k$ and $\partial^*_k$ 
on a function on $\C^D$ as
$(\partial/\partial x_k \mp i \partial/\partial y_k)/2$,
respectively and let $\phi(z)$ be in the class of distributions on $\C^D$
defined above but dropping the restriction $\int\phi(z)\dd^{2D}z \not= 0$.
Then if $p(z)$ is a probability, $p(z)+\partial^*_k\phi(z)$ is also a
probability and in fact (unitarily) equivalent to $p(z)$,
\begin{equation}
\int Q(z) \partial^*_k\phi(z)\dd^{2D}z = 
\int \partial^*_k(Q(z)\phi(z))\dd^{2D}z = 0 \,,
\label{eq3}
\end{equation}
where $Q(z)$ is any analytical polynomial.
That is, $\partial^*_k\phi(z)$ would represent the zero distribution
on $\R^D$. Such distributions will be called null distributions. They
will prove useful in what follows to obtain positive representations
from real ones, namely, by adding null distribution of the form
$\sum_{k=1}^D\partial_k\partial^*_k\phi_k(z)$, for suitably chosen
real $\phi_k(z)$. Note that $4\partial_k\partial^*_k$ is just a Laplacian.

Similarly, by proceeding as in eq.~(\ref{eq3}), it follows that
if $p(z)$ represents $P(x)$, the following relations hold
\begin{eqnarray}
\int Q(z) \partial_k p(z)\dd^{2D}z &=& \int Q(x) \partial_k P(x)\dd^Dx
\nonumber \\
\int Q(z) R(z) p(z)\dd^{2D}z &=& \int Q(x) R(x) P(x)\dd^Dx
\end{eqnarray}
where $Q(x)$ and $R(x)$ are arbitrary polynomials.
That is, $\partial_k$ on $\C^D$ represents $\partial_k$ on
$\R^D$ and multiplication by an analytical polynomial $R(z)$
represents multiplication by $R(x)$.

Another interesting construction is related to convolutions. 
The convolution exist for any two complex probabilities since
it can be defined through the product of their Fourier transforms which are
regular distributions.
If $p_1(z)$ and $p_2(z)$ are unitary representations of $P_1(x)$
and $P_2(x)$ respectively, their convolution $p_1*p_2$ is a
unitary representation of $P_1*P_2$. Indeed, $p_1\otimes p_2$ is a
unitary representation of $P_1\otimes P_2$ and
\begin{eqnarray}
\langle z_{i_1}\cdots z_{i_n}\rangle_{p_1*p_2} &=&
\langle (z^{(1)}_{i_1}+z^{(2)}_{i_1})\cdots
(z^{(1)}_{i_n}+z^{(2)}_{i_n})\rangle_{p_1\otimes p_2} \nonumber \\
&=& \langle (x^{(1)}_{i_1}+x^{(2)}_{i_1})\cdots
(x^{(1)}_{i_n}+x^{(2)}_{i_n})\rangle_{P_1\otimes P_2} =
\langle x_{i_1}\cdots x_{i_n}\rangle_{P_1*P_2} \,.
\label{eq4} 
\end{eqnarray}
Furthermore, if $p_1(z)$ and $p_2(z)$ are positive, $p_1*p_2$ is
positive too.
In particular, this allows for obtaining equivalent representations of
known ones: if $p(z)$ is a unitary representation of $P(x)$ and $C(z)$
is a unitary representation of $\delta(x)$, the $D$-dimensional Dirac
delta function, $p*C$ will be unitarily equivalent to $p(z)$, since
$P*\delta=P$. Any probability $C(z)$ normalized to one
defines a unitary representation of $\delta(x)$ if it is
invariant under global phase rotations, i.e., $C(e^{i\varphi}z)=C(z)$
for any $\varphi\in\R$. In this case
\begin{equation}
\int z_{i_1}\cdots z_{i_n} C(z)\dd^{2D}z = \delta_{n,0} \,,
\label{eq5}
\end{equation}
since the angular average of $z_{i_1}\cdots z_{i_n}$ vanishes for
$n>0$. In fact this construction can be regarded as adding a Laplacian,
namely, $p*C-p$, as it is easily seen after Fourier transform. This
procedure can be used to obtain positive representations from real
ones. On the other hand, it  shows that if a complex probability admits a
unitary positive representation it is not unique.

A unitary representation can always be obtained for any $P(x)$ by
taking $p(z)=P(x)\delta(y)$. If $P(x)$ is positive so will be $p(z)$.
This can be generalized as follows. Let $P_0(x)$ be positive
and $P(x)=P_0(x-it)$, $t\in\R^D$ (i.e., a complex translation
under the conditions considered above for affine transformations). Then
$p(z)=P_0(x)\delta(y-t)$ is a unitary positive representation of
$P(x)$. If we allow $P_0$ to depend on $t$, taking linear
combinations we obtain that $p(z)=p(x,y)$
is a unitary representation of
\begin{equation}
P(x)=\int p(x-iy,y)\dd^Dy\,.
\label{eq6}
\end{equation}
This relation has been noted before in the
literature~\cite{OS91,Sa93}, considered as a projection from
probabilities on $\C^D$ to probabilities on $\R^D$.
Note, however, that when this relation can be applied it gives just
one of the $P(x)$ represented by $p(z)$. In fact, since the momenta of
$P(x)$ are the Taylor expansion coefficients of its Fourier transform,
there are many complex probabilities characterized by the same
momenta. As we have seen,
under this projection, the operation $\partial^*_i$ is mapped to
zero. Similarly, $\partial_i$ is mapped to $\partial/\partial x_i$,
and multiplication by an analytical polynomial $Q(z)$ is mapped to
multiplication by $Q(x)$.

As an immediate application of eq.~(\ref{eq6}), we find that for
$m_{ij}$ real, symmetric and positive definite, the probability
\begin{equation}
P(x)= {\tilde f}(x)\exp(-\frac{1}{2}m_{ij}x_ix_j)
\end{equation}
is represented by
\begin{equation}
p(z)=  \det(m)f(my)\exp(-\frac{1}{2}m_{ij}z_iz^*_j) \,,
\label{eq6.1}
\end{equation}
where ${\tilde f}$ is the Fourier transform of $f$ (the repeated
index convention will be used in what follows). For example, for
$D=1$, and $\Gamma$ positive, $P(x)=\cos(x)\exp(-x^2/2\Gamma)$ is
represented by the positive probability $p(z)=
\exp(-x^2/2\Gamma)(\delta(y-\Gamma)+\delta(y+\Gamma))$. Since in this
example $P(x)$ is real but not positive definite, this is an instance
where a complex Langevin simulation would fail~\cite{FO86,FO94,Sa93},
yet there is a positive representation.

Next, let us show that every complex probability on $\R^D$ admits a
real representation. Let $P(x)$ be a complex probability normalized to
one and ${\tilde P}(k)$ its Fourier transform
\begin{equation}
{\tilde P}(k) = \int e^{ikx}P(x)\dd^Dx
\label{eq7}
\end{equation}
where $kx=k_ix_i$. By definition we have
\begin{equation}
{\tilde P}(k) = \sum_{n=0}^\infty
\frac{i^n}{n!} k_{i_1}\cdots k_{i_n }\langle  x_{i_1}\cdots
x_{i_n}\rangle_P
\label{eq8}
\end{equation}
in a neighborhood of $k=0$ since ${\tilde P}(k)$ is analytic at the
origin. Also,
\begin{equation}
\langle  x_{i_1}\cdots x_{i_n}\rangle_P =
(-i)^n \partial_{i_1}\cdots \partial_{i_n}
{\tilde P}(k)\big|_{k=0} \,.
\label{eq9}
\end{equation}
For a probability $p(z)$ on $\C^D$, the Fourier transform is defined
similarly,
\begin{equation}
{\tilde p}(\sigma) = \int e^{ikx+iry}p(z)\dd^{2D}z =
\int e^{i(\sigma^*z+\sigma z^*)/2}p(z)\dd^{2D}z \,,
\label{eq10}
\end{equation}
where $\sigma_i=k_i+ir_i$. Assuming that $p(z)$ is normalized to one,
its momenta are obtained through
\begin{equation}
\langle  z_{i_1}\cdots z_{i_n}z^*_{j_1}\cdots z^*_{j_m}\rangle_p =
(-2i)^{n+m} \partial^*_{i_1}\cdots\partial^*_{i_n}
\partial_{j_1}\cdots\partial_{j_m}
{\tilde p}(\sigma)\big|_{\sigma=0}
\label{eq11}
\end{equation}
where $\partial_i$ refers to $\sigma_i$ and $\partial^*_j$ to
$\sigma^*_j$.
Consider the following probability,
\begin{equation}
{\tilde p}(\sigma) = {\tilde C}(\sigma){\tilde P}(\frac{\sigma^*}{2})
({\tilde P}(-\frac{\sigma^*}{2}))^*\,.
\label{eq12}
\end{equation}
Here $C(z)$ is one of the real unitary representations of $\delta(x)$
above mentioned. Thus ${\tilde C}(\sigma)$ is analytical at the origin
as a function of $k_i$ and $r_i$ and is invariant under global phase
rotations of $\sigma$.  ${\tilde P}(\sigma)$ stands for the analytical
extension of ${\tilde P}(k)$ in a neighborhood of the origin. Beyond
the analyticity circle (if it is finite) we can choose ${\tilde
C}(\sigma)$ equal to zero so that ${\tilde p}(\sigma)$ exists.  By
construction, ${\tilde p}(\sigma)$ is unity at the origin and
analytical there. Also it is locally integrable and, with a suitable
choice of ${\tilde C}(\sigma)$, grows at most polynomically at
infinity, therefore it defines a probability $p(z)$ on
$\C^D$. Furthermore, $p(z)$ is real since $C(z)$ is real and $({\tilde
p}(\sigma))^*={\tilde p}(-\sigma)$.  It remains to show that it is a
representation of $P(x)$,
\begin{eqnarray}
\langle  z_{i_1}\cdots z_{i_n}\rangle_{p} &=&
(-2i)^n \partial^*_{i_1}\cdots\partial^*_{i_n}
{\tilde p}(\sigma)\Big|_{\sigma=0} = 
(-2i)^n \partial^*_{i_1}\cdots\partial^*_{i_n}
{\tilde P}(\frac{\sigma^*}{2})\Big|_{\sigma=0} \nonumber \\
&=&
(-i)^n\partial_{i_1}\cdots \partial_{i_n}
{\tilde P}(k)\Big|_{k=0}
= \langle  x_{i_1}\cdots x_{i_n}\rangle_P \,,
\label{eq13}
\end{eqnarray}
where it has been used that
$\partial^*_{i_1}\cdots\partial^*_{i_n}{\tilde
C}(\sigma)\big|_{\sigma=0}$ vanishes for $n>0$. That is, we have given
a constructive method, eq.~(\ref{eq12}), to obtain a real
representation of any complex probability within the class of complex
probabilities considered.

As an illustration, consider $D=1$ and 
\begin{equation}
P(x)=\delta(x)+a\delta^\prime(x)\,,\qquad a=a_R +i a_I \in \C \,.
\label{eq14}
\end{equation}
In this case ${\tilde P}(\sigma)= 1-ia\sigma$ is a polynomial, thus it
is entire 
and well-behaved at infinity and we can take ${\tilde C}(\sigma)=1$,
i.e., $C(z)=\delta(x)\delta(y)$. With this choice
\begin{equation}
{\tilde p}(\sigma)=
1-\frac{1}{4}|a|^2|\sigma|^2-\frac{i}{2}(a^*\sigma+a\sigma^*)
\label{eq15}
\end{equation}
and
\begin{equation}
p(z)= \delta(x)\delta(y)
+a_R\delta^\prime(x)\delta(y)+a_I\delta(x)\delta^\prime(y) 
+\frac{1}{4}|a|^2(\delta^{\prime\prime}(x)\delta(y)+
\delta(x)\delta^{\prime\prime}(y)) \,.
\label{eq16}
\end{equation}
One can easily check that this is a real distribution which represents
$P(x)$, however it is not positive. We can find a positive
representation by first applying a convolution (i.e., a better choice
of $C(z)$) and then adding a suitable Laplacian. Furthermore, it can
be done for an arbitrary distribution of support at zero in any number
of dimensions.  Rather than showing this in detail here, it will be
obtained as a byproduct in the next section. There we will obtain
positive representations of Gaussian functions times polynomials.

By formally undoing the Fourier transform of ${\tilde p}(\sigma)$ in
eq.~(\ref{eq12}), the following explicit form of $p(z)$ is obtained
\begin{equation}
p(z) = \int
C_0(x-\frac{x_1+x_2}{2},y-\frac{x_1-x_2}{2i})P(x_1)P^*(x_2)\dd^Dx_1
\dd^Dx_2 \,,
\label{eq16.1}
\end{equation}
where $C_0(z_1,z_2)$ is the analytical extension of $C_0(x,y) =
C(x+iy)$, with $x$ and $y$ real. In order for this formula to make
sense, we should require $C_0(z_1,z_2)$ to be entire on $\C^{2D}$ and
further the integrand should be sufficiently convergent so as to define a
probability on $\C^D$. Such probability is real by construction, since
$C(z)$ is real, however it will not be positive in general even if
$C(z)$ is positive since such property is lost after analytical
extension. The interest of this relation, as compared, for
instance with that in eq.~(\ref{eq6}), is that it is constructive.

An example of application of this formula is provided by
\begin{equation}
P(x)= \sum_{i=1}^Na_i\delta(x-x^{(i)}), \qquad C(z)=
\exp(-\frac{z_jz^*_j}{2\Gamma}) \,,
\end{equation}
which gives
\begin{equation}
p(z) = \sum_{i,j=1}^N a_ia^*_j\exp\left(-\frac{1}{2\Gamma}\left(
\left(x-\frac{x^{(i)}+x^{(j)}}{2}\right)^2
+ \left(y-\frac{x^{(i)}-x^{(j)}}{2i}\right)^2\right)\right)
\,.
\end{equation}
Another application is when $P(x)$ is a finite linear combination of
Gaussian distributions centered anywhere in the complex plane and with
arbitrary complex widths, provided we choose $\Gamma > |\Gamma_i|$,
$i=1,\dots,N$.

\section{Positive representations of Gaussian distributions}

A Gaussian complex probability takes the general form
\begin{eqnarray}
G(x) &=& N_G\exp(-\frac{1}{2}m_{ij}x_ix_j-b_ix_i)\,, \nonumber \\
N_G &=& (2\pi)^{-D/2}(\det(m))^{1/2}\exp(-\frac{1}{2}(m^{-1})_{ij}b_ib_j)
\label{eq17}
\end{eqnarray}
where $m_{ij}$ is a symmetric complex matrix with positive definite
real part to ensure normalizability. As a consequence $m_{ij}$ is non
singular and can be written as $A_{ki}A_{kj}$. This allows to
set $m_{ij}=\delta_{ij}$ and $b_i=0$ by means of a complex affine
transformation. That is, we will consider only
\begin{equation}
G(x)=(2\pi)^{-D/2}\exp(-\frac{1}{2}x_ix_i)
\end{equation}
and the general case can be obtained a posteriori as $G(Ax+A^{-1}b)$.
A positive representation of $G(x)$ is simply
$G(x)\delta(y)$. A more general representation $g(z)$ is obtained by
convolution with $C(z)=(2\pi\eta)^{-D}\exp(-\frac{1}{2\eta} z_iz^*_i)$,
where $\eta$ is positive. This gives
\begin{eqnarray}
g(z) &=& N_g\exp(-\frac{1}{2(\eta+1)}x_ix_i-\frac{1}{2\eta}y_iy_i) \nonumber\\
&=&  N_g\exp(-\bar\gamma z_iz^*_i +\frac{1}{2}\gamma z_iz_i
+\frac{1}{2}\gamma z^*_iz^*_i)\,,
\label{eq18}
\end{eqnarray}
where the normalization constant is 
$N_g=\left(2\pi\sqrt{\eta(\eta+1)}\right)^{-D}$ and we have introduced the
positive numbers
\begin{equation}
\gamma= \frac{1}{4\eta(\eta+1)}\,,\qquad
\bar\gamma=\frac{2\eta+1}{4\eta(\eta+1)}\,.
\label{eq19}
\end{equation}
The same representation is obtained by following the method of
eq.~(\ref{eq16.1}). The value of the parameter $\eta$, or equivalently
$\bar\gamma$, will be fixed below.

The set of probabilities to be considered is $P(x)=Q(x)G(x)$, where
$Q(x)$ is a complex polynomial of degree $N$. $P(x)$ can always be written as
\begin{equation}
P(x)= \sum_{n=0}^N\frac{1}{n!}
a_{i_1\dots i_n}\partial_{i_1}\dots \partial_{i_n}G(x)\,,
\label{eq20}
\end{equation}
where $a_{i_1\dots i_n}$ is completely symmetric and the zeroth order
coefficient $a_0$ must not vanish (in fact, is unity if $P(x)$
is normalized).
A real representation of $P(x)$ is given by
\begin{equation}
p_0(z)= (|a_0|^2+
a_0^*\sum_{n=1}^N\frac{1}{n!}
a_{i_1\dots i_n}\partial_{i_1}\dots \partial_{i_n} +
a_0\sum_{n=1}^N\frac{1}{n!}
a^*_{i_1\dots i_n}\partial^*_{i_1}\dots \partial^*_{i_n})g(z)\,,
\label{eq21}
\end{equation}
since the terms with $\partial^*$ do not contribute and
$\partial/\partial z$ is  mapped to $\partial/\partial x$ under projection.

It is convenient to introduce the polynomials
\begin{equation}
Q_{i_1\dots i_n}(z) = g(z)^{-1}\partial_{i_1}\cdots\partial_{i_n}g(z)\,.
\label{eq22}
\end{equation}
They can be computed recursively by means of the formula
\begin{equation}
Q_0(z) = 1\,,\qquad
Q_{i_1\dots i_n}(z) = (\partial_{i_n}+\omega_{i_n})Q_{i_1\dots i_{n-1}}(z)\,,
\label{eq22.1}
\end{equation}
where we have introduced the variable
\begin{equation}
\omega_i = \gamma z_i-\bar\gamma z^*_i \,.
\label{eq22.2}
\end{equation}
The functions $Q_{i_1\dots i_n}(z)$ are polynomials of degree $n$ in
$\omega_i$, with coefficients depending only on $\gamma$.
With this notation, $p_0(z)$ can be rewritten as
\begin{equation}
p_0(z)= \left(|a_0|^2+ a_0^*\sum_{n=1}^N\frac{1}{n!}
a_{i_1\dots i_n}Q_{i_1\dots i_n}(z) +
a_0\sum_{n=1}^N\frac{1}{n!}a^*_{i_1\dots i_n}Q^*_{i_1\dots i_n}(z)\right)
g(z)\,.
\label{eq23}
\end{equation}
In order to obtain a positive representation, $p(z)$ can be further
cast in the form
\begin{eqnarray}
p_0(z) = \Big(|a_0|^2
&+& \sum_{n=1}^N\frac{1}{n!}\beta_n
|Q_{i_1\dots i_n}(z)+\beta_n^{-1}a_0a^*_{i_1\dots i_n}|^2  \nonumber \\
&-&\sum_{n=1}^N
(\frac{1}{n!}\beta_n|Q_{i_1\dots i_n}(z)|^2+\beta_n^{-1}|a_0|^2|a_n|^2
)\Big)g(z)\,,
\label{eq24}
\end{eqnarray}
where the indices $i_1\dots i_n$ are summed over.
$\beta_1,\dots,\beta_N$ are arbitrary positive numbers which value is to be 
specified below and we have
defined the quantity $|a_n|$ as
\begin{equation}
|a_n|^2=\frac{1}{n!}a_{i_1\dots i_n}a^*_{i_1\dots i_n}\,.
\end{equation}
We will assume that $|a_n|$ is non vanishing, since the vanishing
case is trivial.
In Appendix A it is shown that
\begin{equation}
\phi_n(z) = \left(\frac{1}{n!}Q_{i_1\dots i_n}(z)Q^*_{i_1\dots i_n}(z)
 -{\bar\gamma}^n K_n(D)\right)g(z)
\label{eq25}
\end{equation}
is a null distribution, where
\begin{equation}
K_n(D) = \frac{(D+n-1)!}{n!(D-1)!}\,.
\end{equation}
By removing $\phi_n(z)$ from $p_0(z)$ we obtain an equivalent
representation $p(z)$, namely, 
\begin{equation}
p(z) =  \left[\sum_{n=1}^N\frac{1}{n!}
\beta_n|Q_{i_1\dots i_n}(z)+\beta_n^{-1}a_0a^*_{i_1\dots i_n}|^2
+|a_0|^2-\sum_{n=1}^N(\beta_n{\bar\gamma}^n K_n(D)+
\beta_n^{-1}|a_0|^2|a_n|^2)\right]g(z) \,.
\label{eq26}
\end{equation}
To ensure positivity of $p(z)$ we require
\begin{equation}
\sum_{n=1}^N(\beta_n{\bar\gamma}^n K_n(D)+
\beta_n^{-1}|a_0|^2|a_n|^2) \leq |a_0|^2\,.
\label{eq27}
\end{equation}
This can be achieved by choosing the positive coefficients 
$\beta_n$ so as to minimize the left-hand side, 
\begin{equation}
\beta_n = \frac{|a_0||a_n|}{\sqrt{\bar\gamma^nK_n(D)}}\,.
\label{eq28}\end{equation}
In this way the inequality is satisfied for any
$\bar\gamma$ smaller than the unique positive solution of 
\begin{equation}
\sum_{n=1}^N \sqrt{K_n(D)}|a_n|{\bar\gamma}^{n/2}
= \frac{1}{2}|a_0| \,.
\label{eq29}
\end{equation}
For this choice of $\bar\gamma$, $p(z)$ takes the simple form
\begin{equation}
p(z) = \sum_{n=1}^N\frac{1}{n!}
\beta_n|Q_{i_1\dots i_n}(z)+\beta_n^{-1}a_0a^*_{i_1\dots i_n}|^2
g(z)\,.
\label{eq26.1}
\end{equation}
To summarize, any Gaussian times polynomial complex probability,
eq.~(\ref{eq20}), admits a positive representation, namely, $p(z)$ in 
eq.~(\ref{eq26.1}), with $\beta_n$ 
given by eq.~(\ref{eq28}), and $\bar\gamma$ given by
eq.~(\ref{eq29}).

Incidentally, let us note that from a computational point of view, it
is convenient to minimize the width of $p(z)$ in the complex plane
(e.g., if $P(x)$ is already positive, the best choice is
$P(x)\delta(y)$), since this reduces the dispersion of points in the
sample. In the family of probabilities described by the expression of
$p(z)$ in eq.~(\ref{eq26}), this minimization corresponds to our
choice of $\beta_n$ in eq.~(\ref{eq28}) and $\bar\gamma$ in
eq.~(\ref{eq29}). In general, however, this needs not be best
equivalent positive representation of $P(x)$. The construction
presented above corresponds to adding to $p_0(z)$ a Laplacian of the
form $\partial_{i_1}\cdots\partial_{i_n}
\partial^*_{i_1}\cdots\partial^*_{i_n}g(z)$ (as can be seen using the
formulas of Appendix A). More generally, one could add terms of the
form $b_{i_1\dots i_n;j_1\dots j_n} \partial_{i_1}\cdots\partial_{i_n}
\partial^*_{j_1}\cdots\partial^*_{j_n}g(z)$, with $b$ self-adjoint, in
order to optimize $p(z)$, or even more general terms so long as they
have a $\partial^*_j$ and are real.

Let us now come back to the problem of finding positive
representations of complex distributions with support at {0}. Such
distributions take the form
\begin{equation}
P(x)= \sum_{n=0}^N\frac{1}{n!}
a_{i_1\dots i_n}\partial_{i_1}\dots \partial_{i_n}\delta(x)\,.
\label{eq30}
\end{equation}
This distribution can be considered as the zero width limit of the
Gaussian times polynomial distribution. 
\begin{equation}
P(x)= \lim_{\lambda\to 0^+}P_\lambda(x)\,,\quad
P_\lambda(x)= \sum_{n=0}^N\frac{1}{n!}
a_{i_1\dots i_n}\partial_{i_1}\dots \partial_{i_n}(\lambda^{-D}G(x/\lambda))\,.
\end{equation}
Naming $P(x;a)$ the probability in eq.~(\ref{eq20}), we find 
\begin{equation}
P_\lambda(x)= \lambda^{-D}P(x/\lambda;a^\lambda)\,,\qquad a_{i_1\dots
i_n}^\lambda = \lambda^{-n} a_{i_1\dots i_n} \,.
\end{equation}
Therefore, the positive representation of $P(x;a)$, namely, $p(z;a)$
in eq.~(\ref{eq26.1}), provides a positive representation of
$P_\lambda(x)$,
\begin{equation}
p_\lambda(z)=
\lambda^{-2D}p(z/\lambda;a^\lambda) \,.
\end{equation}
In order to take the limit, we should consider how the different
variables scale. We already have the scaling law of $z$ and of the
coefficients $a_{i_1\dots i_n}$. From eqs.~(\ref{eq28},\ref{eq29})
$\beta^\lambda_n$ is found to scale as $\lambda^{-2n}\beta_n$ and
$\bar\gamma^\lambda$ as $\lambda^2\bar\gamma$. From eqs.~(\ref{eq19}),
$\eta^\lambda$ is given in leading order by $\lambda^{-2}\eta$ with
$\eta=1/(2\bar\gamma)$ and $\gamma^\lambda$ is of order $\lambda^4$ and
can be neglected. Therefore, in leading order
$\lambda^{-2D}g(z/\lambda;\bar\gamma^\lambda)$ becomes
\begin{equation}
g^0(z;\bar\gamma)= (2\pi\eta)^{-D}\exp(-\bar\gamma z_iz^*_i)\,,\quad
\eta=\frac{1}{2\bar\gamma}
\label{eq31}
\end{equation}
and is independent of $\lambda$. This results is to be used in
eq.~(\ref{eq26.1}).
Finally, in leading order, $Q_{i_1\dots i_n}(z/\lambda;\bar\gamma^\lambda)$
becomes $\lambda^n Q^0_{i_1\dots i_n}(z;\bar\gamma)$ with
\begin{equation}
Q^0_{i_1\dots i_n}(z;\bar\gamma)= 
g^0(z;\bar\gamma)^{-1}\partial_{i_1}\cdots\partial_{i_n}g^0(z;\bar\gamma) =
(-\bar\gamma)^n z^*_{i_1}\cdots z^*_{i_n} \,.
\label{eq32}
\end{equation}
To summarize, any complex distribution with support at a single point,
eq.~(\ref{eq30}),
admits a positive representation, namely, 
\begin{equation}
p(z) = \sum_{n=1}^N\frac{1}{n!}
\beta_n|Q^0_{i_1\dots i_n}(z)+\beta_n^{-1}a_0a^*_{i_1\dots i_n}|^2
g^0(z)\,.
\end{equation}
with $\beta_n$ given by eq.~(\ref{eq28}), and $\bar\gamma$ given by
eq.~(\ref{eq29}).

As an illustration we can consider again the distribution of
eq.~(\ref{eq14}). In this case we find $\bar\gamma = (4|a|^2)^{-1}$
and $\eta = \beta_1= 2|a|^2$, and thus
\begin{equation}
p(z) = |z-2a|^2\exp(-\frac{|z|^2}{4|a|^2})\,.
\label{eq33}
\end{equation}

As a final application of the results of this section, we can consider
periodic probabilities.
Such probabilities correspond to variables effectively defined in a
compact domain and find application in the context of compact gauge
theories on the lattice. 
They satisfy, $P(x)= P(x-na)$ with $(na)_i=n_ia_i$
where $n\in\Z^D$ is arbitrary and $a\in\R_+^D$ is characteristic of
$P(x)$. Without loss of generality, we may choose $a_i=2\pi$.
These probabilities do not belong to the class previously
considered. The normalization as well as the expectation values
should be taken on a lattice cell $\{x, 0\le x_i<2\pi, i=1,\dots,D\}$.
The test functions should be periodic and 
the concept of representation should be modified accordingly: $p(z)$
is periodic on the real axis, $x$ is to be integrated on the periodic cell
and $y$ on $\R^D$. Also
instead of equality of expectation values of polynomials we demand
$\langle\exp(in_jx_j)\rangle_P
=\langle\exp(in_jz_j)\rangle_p$ for any integers $n_j$, $j=1,\dots,D$.
Assume now that the periodic distribution is a function of the form
\begin{equation}
P(x)= \sum_{n\in\Z^D}P_0(x-2\pi n),
\end{equation}
where the series is uniformly convergent. Let $p_0(z)$ be a function
which is a
positive representation of $P_0(x)$ not only on polynomials but also
on exponential test functions, and such that
\begin{equation}
p(z)= \sum_{n\in\Z^D}p_0(z-2\pi n) \,.
\end{equation}
is uniformly convergent. Then, $p(z)$ is a positive 
representation of $P(x)$, as is readily shown.

In particular, $P_0(x)$ may be a Gaussian times polynomial and
$p_0(z)$ its positive representation found above, since these functions are
sufficiently convergent at infinity. Therefore
the construction given above provides a positive representation for
this case too. Another example is the periodic version of the one
dimensional Gaussian times cosine considered above after eq.~(\ref{eq6.1}):
\begin{eqnarray}
P(x) &=&\cos(x)\sum_{n\in\Z}\exp\left(-\frac{(x-2\pi
n)^2}{2\Gamma}\right) \,, \nonumber \\
p(z) &=&(\delta(y-\Gamma)+\delta(y+\Gamma))
\sum_{n\in\Z}\exp\left(-\frac{(x-2\pi n)^2}{2\Gamma}\right) \,.
\end{eqnarray}
This example is interesting since it is similar to simplified
probabilities considered in the literature~\cite{FO86,FO94} to model
the SU(2) gauge theory in the presence of a Wilson loop, for which the
complex Langevin algorithm did not work.

\section{Concluding remarks}

We have studied the problem of representation of complex distributions
by distributions on the analytically extended complex plane. The
positive representation problem is of immediate interest in some areas
of physics: field theory and statistical mechanics. On the other hand
it also seems a new and interesting field from the mathematical point
of view. One could consider extending the particular class of complex
distributions studied here, namely, Fourier transforms of regular
distributions analytical at the origin, by allowing as well for adding
non regular distributions with support outside the origin.  Perhaps
more interesting, and in the opposite direction, one could extend the
set of test functions in the definition of representation beyond
polynomials to insure, for instance, that each probability on $\C^D$
is at most the representation of one probability on $\R^D$. From the
viewpoint of applications it would also be interesting to extend the
concept of representations to distributions defined on group manifolds
since they appear naturally in lattice gauge theories. Our discussion
on periodic distributions corresponds in fact to the manifold of the
direct product of $D$ U(1) factors.

\section{Acknowledgments}
I would like to thank C. Garc\'{\i}a-Recio for comments on the
manuscript.

\appendix{}\section{}

In this appendix we will show that $\phi_n(z)$ defined in
eq.~(\ref{eq25}) is a null distribution. To this end let us introduce
the polynomials
\begin{equation}
Q_{i_1\dots i_n;j_1\dots j_m}(z) =
g(z)^{-1}\partial_{i_1}\cdots\partial_{i_n} 
\partial^*_{j_1}\dots\partial^*_{j_m}g(z) \,.
\label{eqA1}
\end{equation}
They generalize $Q_{i_1\dots i_n}(z)$ and satisfy the relation
\begin{equation}
Q_{i_1\dots i_n;j_1\dots j_m}(z)= Q^*_{j_1\dots j_m;i_1\dots i_n}(z)\,.
\label{eqA2}
\end{equation}
To prove eq.~(\ref{eq25}), we will use the following Wick theorem:
\begin{equation}
Q_{i_1\dots i_n}(z)Q^*_{j_1\dots j_m}(z)= \sum_{[i_1\dots i_n;j_1\dots j_m]}
Q_{i_1\dots i_n;j_1\dots j_m}(z)\,.
\label{eqA3}
\end{equation}
where the sum is over all possible sets of contractions of the indices
$i_1\dots i_n$ with the indices $j_1\dots j_m$. The contraction of two
indices $i$, $j$ gives a factor $\bar\gamma\delta_{ij}$ and removes
them from the list, e.g.,
\begin{equation}
Q_{i_1i_2}(z)Q^*_j(z)= Q_{i_1i_2;j}(z) + \bar\gamma\delta_{i_1j}Q_{i_2}(z) +
\bar\gamma\delta_{i_2j}Q_{i_1}(z) \,.
\label{eqA4}
\end{equation}
In general there are $n!m!/k!(n-k)!(m-k)!$ terms with $k$
contractions. Let us apply the Wick theorem
to $Q_{i_1\dots i_n}(z)Q^*_{j_1\dots j_n}(z)g(z)$. Whenever two indices $i,j$
are not contracted we will have $Q_{i\dots;j\dots}(z)g(z)$ which
contains $\partial^*_j$ and hence is a null distribution. Therefore
only the terms with all indices contracted contribute and the non
null part is
\begin{equation}
\bar\gamma^n\sum_{p\in S_n}\delta_{i_1j_{p1}}\cdots\delta_{i_nj_{pn}}g(z)\,,
\label{eqA5}
\end{equation}
where the sum runs over all permutations. After contracting the indices we
obtain eq.~(\ref{eq25}). $K_n(D)$ is the number of ways of choosing $n$
objects out of $D$ allowing repetitions.

The Wick theorem can be proven by induction. Defining the operator
\begin{equation}
{\cal D}_i = g^{-1}(z)\partial_i g(z) = \partial_i+\omega_i \,,
\end{equation}
($g(z)$ is a multiplicative operator here) we have
\begin{eqnarray}
Q_{i_1\dots i_n}(z) &=& {\cal D}_{i_1}\cdots{\cal D}_{i_n}Q_0(z) \,,
\nonumber \\
Q_{i_1\dots i_n;j_1\dots j_m}(z) &=& {\cal D}_{i_1}\cdots{\cal
D}_{i_n}{\cal D}^*_{j_1}\cdots{\cal D}^*_{j_n}Q_0(z) \,,
\label{eqA6}
\end{eqnarray}
where $Q_0(z)=1$. Trivially, $[\partial_i,{\cal D}^*_j] =
-\bar\gamma\delta_{ij}$, thus
\begin{equation}
\partial_i Q^*_{j_1\dots j_m}(z) =
-\sum_{k=1}^m\bar\gamma\delta_{ij_k}Q^*_{j_1\dots \hat{j_k}\dots j_m}(z)\,,
\label{eqA7}
\end{equation}
where the hat means that the index has been removed from the list.
On the other hand ${\cal D}(AB)=({\cal D}A)B-A\partial B$. The Wick theorem
holds for $n=m=0$. Assuming it has been proven up to some
$(n,m)$,
\begin{eqnarray}
Q_{i_1\dots i_{n+1}}(z)Q^*_{j_1\dots j_m}(z) &=&
({\cal D}_{i_{n+1}}Q_{i_1\dots i_n}(z))Q^*_{j_1\dots j_m}(z) \nonumber \\
&=& {\cal D}_{i_{n+1}}(Q_{i_1\dots i_n}(z)Q^*_{j_1\dots j_m}(z))
- Q_{i_1\dots i_n}(z)\partial_{i_{n+1}}Q^*_{j_1\dots j_m}(z) \,.
\end{eqnarray}
Using that the theorem holds for $(n,m)$ and eq.~(\ref{eqA7}),
\begin{eqnarray}
Q_{i_1\dots i_{n+1}}(z)Q^*_{j_1\dots j_m}(z) &=&
\sum_{[i_1\dots i_n;j_1\dots j_m]} Q_{i_1\dots i_{n+1};j_1\dots
j_m}(z) \nonumber \\
&+& 
\sum_{k=1}^m\sum_{[i_1\dots i_n;j_1\dots \hat{j_k}\dots j_m]} \bar\gamma
\delta_{i_{n+1}j_k}Q_{i_1\dots i_n;j_1\dots \hat{j_k}\dots j_m}(z) \,.
\end{eqnarray}
The first term contains all the contractions not involving the index
$i_{n+1}$, and the second one all the contractions involving the index
$i_{n+1}$, hence the theorem is proven for $(n+1,m)$. It is
worth noticing that the reverse expansion also holds, i.e.,
\begin{equation}
Q_{i_1\dots i_n;j_1\dots j_m}(z)= \sum_{[i_1\dots i_n;j_1\dots j_m]}
Q_{i_1\dots i_n}(z)Q^*_{j_1\dots j_m}(z)\,.
\end{equation}
where the contraction of $ij$ now is $-\bar\gamma\delta_{ij}$.

\end{document}